\documentclass[%
 reprint,
 amsmath,amssymb,
 aps,
]{revtex4-2}

\usepackage{graphicx}
\usepackage{dcolumn}
\usepackage{bm}
\usepackage{multirow}
\usepackage{lineno}

\usepackage{placeins}

\usepackage{float}

\usepackage{url}
\usepackage{hyperref}
\usepackage[hyphenbreaks]{breakurl}

\begin{document}


\title{Aircrew rostering workload patterns and associated fatigue and sleepiness scores in short/medium haul flights under RBAC 117 rules in Brazil}

\author{Tulio E. Rodrigues$^{1,2,*}$, Eduardo Furlan$^{2}$, André F. Helene$^{3}$, Otaviano Helene$^{1}$, Eduardo Pessini$^{4}$, Alexandre Simões$^{5}$, Maurício Pontes$^{6}$ and Frida M. Fischer$^{7}$ \vspace{0.15in}}

\affiliation{$^{1}$Experimental Physics Department, Physics Institute, University of São Paulo, P. O. Box 66318, CEP 05315-970, São Paulo, Brazil}
\affiliation{$^{2}$ Technical Board, Gol Aircrew Association (ASAGOL), São Paulo, Brazil}
\affiliation{$^{3}$Department of Physiology, Institute of Biosciences, University of São Paulo, São Paulo, Brazil}
\affiliation{$^{4}$Aeronautics Institute of Technology (ITA), São Paulo, Brazil}
\affiliation{$^{5}$Safety Board, LATAM Aircrew Association (ATL), São Paulo, Brazil}
\affiliation{$^{6}$Technical Board, Brazilian Association of Civil Aviation Pilots (ABRAPAC), São Paulo, Brazil}
\affiliation{$^{7}$ Department of Environmental Health, School of Public Health, University of São Paulo, São Paulo, Brazil}

\date{\today}

\begin{abstract}
The relationships between workload and fatigue or sleepiness are investigated through the analysis of rosters and responses to questionnaires from Brazilian aircrews, taken from \textit{Fadigômetro} database.
The approach includes temporal markers - coinciding with Samn-Perelli (SP) and Karolinska Sleepiness Scale (KSS) responses - where SAFTE-FAST model outcomes are calculated.
The model results follow the increase of fatigue and sleepiness perceptions during the dawn (0h00 to 05h59), but underestimate the self-rated scores during the evening (18h00 to 23h59). On the other hand, the KSS scores fit the relative risk of pilot errors, representing a reasonable proxy for risk assessment.
Linear relationships obtained between workload metrics, computed within 168-hours prior to the responses, and self-rated SP and KSS scores provide a consistent method to estimate accumulated fatigue and sleepiness.
Considering 7149 rosters of 2023, the duty time ($DT$), the number of flight sectors ($N_{CREW}$) and the sum of flight sectors with sit periods longer than one hour ($N_{CREW}+N_{SIT} $) are associated with 70.1\%/60.6\% of the highest predicted scores of SP/KSS. Applying the mitigations $DT\leq 44 h$, $N_{CREW}\leq 15$ and $N_{CREW}+N_{SIT} \leq 19$ for every 168-hour interval yields a significant decrease in the higher values of SP/KSS with minimal impact on aircrew productivity.

\end{abstract}


\email{tulio@if.usp.br}

\maketitle


\section{Introduction}

According to the International Civil Aviation Organization (ICAO) \citep{ICAO2016} fatigue is defined as ``a physiological state of reduced mental or physical performance capability resulting from sleep loss, extended wakefulness, circadian phase, and/or workload (mental and/or physical activity) that can impair a person's alertness and ability to perform safety related operational duties". In this regard, ICAO recommends that operators apply methods and strategies to manage the fatigue risks, either via a Prescriptive Approach, combined with an effective Safety Management System, or via the implementation of a Fatigue Risk Management System \citep{ICAO2016}. 
Despite been necessary in most cultures, the Prescriptive Approach is not sufficient to mitigate fatigue and/or sleepiness in all scenarios, especially during disruptive and/or night shifts, as recently announced by the European Aviation Safety Agency (EASA) \citep{EASA2019}. Such finding makes clear the need for improvements in the Safety Management Systems in order to identify potential hazards and apply effective mitigation barriers, regardless of being fully compliant with the regulations.\\
Both methods of fatigue management should be based on scientific principles and knowledge and operational experience. More specifically, the ``interactions between fatigue and workload in their effects on physical and mental performance" should be taken into account in this framework \citep{ICAO2016}.\\
As pointed out recently \citep{Rod2023}, biomathematical models have been developed since the 80's in order to estimate fatigue and sleepiness levels in industry, but still have their limitations to provide quantitative results. The models usually take into account the sleep-awake cycle, governed by the homeostatic process, and the circadian rhythm oscillations along the day. The effect of sleep inertia – a transient reduction in alertness shortly after waking up – is also considered in some models. Despite being interesting and useful tools for estimating fatigue and/or sleepiness, models may not quantitatively address the complex relationships between task-related workload patterns and fatigue or sleepiness outcomes \citep{Hur2004, olb2011, roma2012, ingre2014, ran2013, raslear2011, bor1982, jewett1999interactive, ake2004, and2004, roach2004model, mor2004, bor2016}.\\ 
Considering the Brazilian regulations, a new set of prescriptive rules was put in place by the local regulator Agência Nacional de Aviação Civil (ANAC) in March of 2020 (Regulamento Brasileiro de Aviação Civil 117, RBAC 117) \citep{Anac2019}, following the requirements of the new Labour Law in effect since August of 2017 \citep{Brasil2017}.\\
In 2017, concomitantly with the new Law, a new collaboration amongst several stakeholders - named \textit{Fadigômetro} - was implemented, with the goal to estimate fatigue and sleepiness outcomes in the analysis of thousands of Brazilian aircrew rosters, as predicted by the Sleep, Activity, Fatigue, and Task Effectiveness Fatigue Avoidance Scheduling Tool (SAFTE-FAST) model \citep{Hur2004}. So far, two studies have been produced under the scope of \textit{Fadigômetro} collaboration, either investigating the sensitivity of SAFTE-FAST outputs when comparing rosters of low- versus high-season months of 2018 \citep{rod2020}, or providing a comprehensive analysis of the root causes of fatigue related with night-shifts and departures and/or arrivals within 02h00 and 06h00 \citep{Rod2023}. This latest study analysed 8476 rosters in a pre-COVID-19 period (2019 and early 2020), proposing important safety recommendations for both the regulatory framework and fatigue risk management policies, which were subsequently approved and endorsed - as preventive measures - by the National Committee for the Prevention of Aeronautical Accidents (CNPAA) in November 2022.\\
Another recent study carried out in Brazil after the implementation of RBAC 117 included 51 airline pilots and demonstrated higher odds for high scores of self-rated fatigue and sleepiness comparing the shifts encompassing 0h00 and 06h00 and early-start shifts (start of duty between 06h01 and 07h59) with the other types of shifts, besides other relevant findings \citep{sampaio2023}.\\
Brazilian regulator ANAC is currently conducting a review of the RBAC 117 rules, proposing considerable increases in several prescriptive limits (including maximum flight duty periods per day, maximum flight hours within 28 days, among others), which makes it necessary that current fatigue levels and associated risks are accurately estimated and appropriately managed. Such premisses depend on a robust and rigorous test of current regulations, which has not yet being accomplished. In this regard, the present work proposes a novel approach to bridge this gap and evaluate the correlations and relationships between several workload patterns in aircrew rosters along each 168-hour loop and the self-rated fatigue and sleepiness scores in terms of the Samn-Perelli \citep{samn1982} and Karolinska Sleepiness Scale \citep{aakerstedt1990}, respectively. The calculations also provide insights into best practices for using bio-mathematical models to guide management decisions, especially when evaluating fatigue and/or sleepiness outcomes in high workload scenarios. Finally, the method also proposes mitigation strategies in some key roster's metrics that are likely to generate small impacts on production during optimization processes with a remarkable reduction in accumulated fatigue and sleepiness.

\section{Methods}

\subsection{The sample and its requirements}
The sample includes the responses of a questionnaire that included socio-demographic, behaviour and health aspects and executed rosters from aircrew workers pertaining to major Brazilian airlines, without distinction of sex, race, age, rank or years in the job. The sample was split into four components:\\
\begin{itemize}
\item Sample 1: 585 aircrew responses of the questionnaire since its launch in June 2021,
\item Sample 2: 263 executed rosters from June 2021 until December 2023 overlapping the questionnaire responses dates for each aircrew,
\item Sample 3: 216 executed rosters from June 2021 until December 2023 overlapping the questionnaire responses dates for each aircrew with valid SAFTE-FAST outputs, and
\item Sample 4: 7149 executed rosters of 2023. 
\end{itemize}
The questionnaire and the overlapping rosters were extracted from the \textit{Fadigômetro} database on January 15, 2024 and all the 2023 rosters were extracted on March 21, 2024.
\subsection{Ethical considerations}
This work was approved by the ethics committee of the Institute of Biosciences, University of São Paulo (Certificate of Presentation for Ethical Appraisal no. 89058318.7.0000.5464). The data acquisition methods and analyses ensure confidentiality to all eligible aircrew volunteers who agreed to participate by approving a digital informed consent form. We declare no conflict of interest with any professional association involved in the experiment, airline or local regulator. The confidentiality of the airlines was also assured.
\subsection{Data acquisition, criteria and markers}
Rosters were acquired through a web application and, more recently, through an application available on IOS and Android mobile platforms. The questionnaire responses were also collected through a web application, where the volunteer accessed his/her exclusive link. Rosters with long-haul flights performed with augmented crew (three or four pilots) within 168 hours prior to the questionnaire response time (hour and minute) were excluded (sample 2), thus ensuring that all rosters were executed with minimal crew (only two pilots).\\
Markers of 15-minutes, ending at the exact questionnaire response time for each aircrew, were merged with the respective roster's files in SAFTE-FAST console. The SAFTE-FAST average Effectiveness and average Sleep Reservoir were then calculated within the markers periods. For details of the SAFTE-FAST model refer to Ref. \citep{Hur2004}. In 47 of the 263 rosters with markers, the SAFTE-FAST model (with all Auto-Sleep functions activated) predicted sleep periods overlapping with questionnaire response times, which prevented valid model outputs. These markers were than discarded in sample 3 (N=216) in order to avoid any arbitrary manipulation of SAFTE-FAST predictive sleep. The SAFTE-FAST parameters and criteria are the same as those adopted in Ref. \citep{Rod2023} and are discussed in Appendix A.\\
Sample 4 included all rosters of 2023 (N=7149), except those with long-haul international flights, thus ensuring that flights were carried out with minimum crew. This eliminates flights with augmented crews that allow in-flight sleep events, which are not considered in the SAFTE-FAST configuration in these analyses. Differently from our previous 30-day epochs for rosters \citep{Rod2023}, this work adopts a 28-day epoch, which is in line with the RBAC 117 parameter for accumulated block hours \citep{Anac2019}. Moreover, this sample also excluded rosters with more than 21 consecutive days without any event, which are likely associated, for instance, with vacation periods encompassing fractions of two consecutive months. As described in more detail elsewhere \citep{Rod2023}, all working and flying activities of the executed rosters were included in the analyses, except home standbys, where the aircrew stays on-call at the place of his/her choice.

\subsection{Fatigue and sleepiness outcomes, workload patterns and statistical analyses}

The model dependent variables include the SAFTE-FAST outputs of Effectiveness [$E_{SF}$(\%)], Sleep Reservoir [$R_{SF}$(\%)], Samn-Perelli Scale ($\text{SP}_{SF}$) and Karolinska Sleepines Scale ($\text{KSS}_{SF}$), all calculated within the 15-minute markers (sample 3). They also include the SAFTE-FAST minimum effectiveness [$EM_{C}$(\%)], minimum sleep reservoir [$RM_{C}$(\%)], and fatigue hazard area [$FHA_{C}$(min)], all within critical phases of flight, considering a standardized 28-days epoch for all rosters of 2023 (sample 4). The self-rated fatigue and sleepiness scores include the Samn-Perelli (SP) \citep{samn1982} and Karolinska Sleepiness Scale (KSS) \citep{aakerstedt1990} from the questionnaire (samples 2 and 3). Given that sample 3 had few results for $\texttt{SP}\geq6$ and $\texttt{KSS}\geq8$, the SP results were grouped from 1 to 5 and a sixth group combining the scores 6 and 7, whereas the KSS scores grouped from 1 to 7 and an eighth group combining the scores 8 and 9. The independent workload metrics were chosen considering the authors' operational expertise and include:
\begin{itemize}
\item Number of night-shifts, $N_{NS}$ (shifts that encompass any fraction between 0h00 and 06h00);
\item Number of consecutive night-shifts, $N_{CNS}$;
\item Number of early-start shifts, $N_{ES}$ (shifts that start between 06h01 and 07h59);
\item Total duty time in hours, $DT(h)$;
\item Number of flight sectors, $N_{CREW}$;
\item Number of departures and/or arrivals between 02h00 and 06h00, $N_{WOCL}$;
\item Number of long sit events, $N_{SIT}$ (intervals between consecutive flight sectors longer than one hour);
\item Number of short rest periods, $N_{REST}$ (rest periods shorter than 16 hours);
\item Number of long duties, $N_{DUTY}$ (duty periods longer than 9 hours). 
\end{itemize}
All these metrics were calculated with a FORTRAN algorithm developed into the Microsoft Visual Studio 2019 version 16.8.5 with Intel® Fortran Compiler – toolkit version: 2021.1.1, considering a time interval of 168 hours (sample 2) or 28 days (sample 4).\\
For the normality hypothesis we have applied the Shapiro-Wilk test \citep{Sha1965}, and for the null hypothesis the Mann-Whitney test for independent samples. All statistical tests, descriptive statistics, calculations of Person's correlations and risk ratios analyses were performed using the IBM SPSS version 25. The linear fittings were done using the Least Squares Method (LSM) in matrix formalism \citep{helene2016useful} with Mathcad 15. The 95\% confidence intervals of the fitted functions were obtained using uncertainty propagation methods and the covariance matrix of the fitted parameters, as described in more detail elsewhere \citep{Rod2023}.

\section{Results}
\subsection{Questionnaire general statistics}
The second questionnaire of \textit{Fadigômetro} study was answered by 585 Brazilian aircrew workers (sample 1) from June 14, 2021 to January 15, 2024. The participants were 69.2\% male, 26.5\% Captains, 22.9\% First Officers and 50.6\% Cabin Crew. The average ages and respective standard deviations in years were: $41.3\pm11.8 $ (male), $37.0	\pm8.6$ (female),  $46.9	\pm12.2$ (Captains), $36.4	\pm8.6$ (First Officers), $38.0	\pm9.8$ (Cabin Crew) and  $42.0\pm11.9$ (Pilots: Captains and First Officers).

\subsection{SAFTE-FAST outputs and self-rated fatigue and sleepiness scores}
\begin{center}
\begin{figure*}
\includegraphics[scale=0.65]{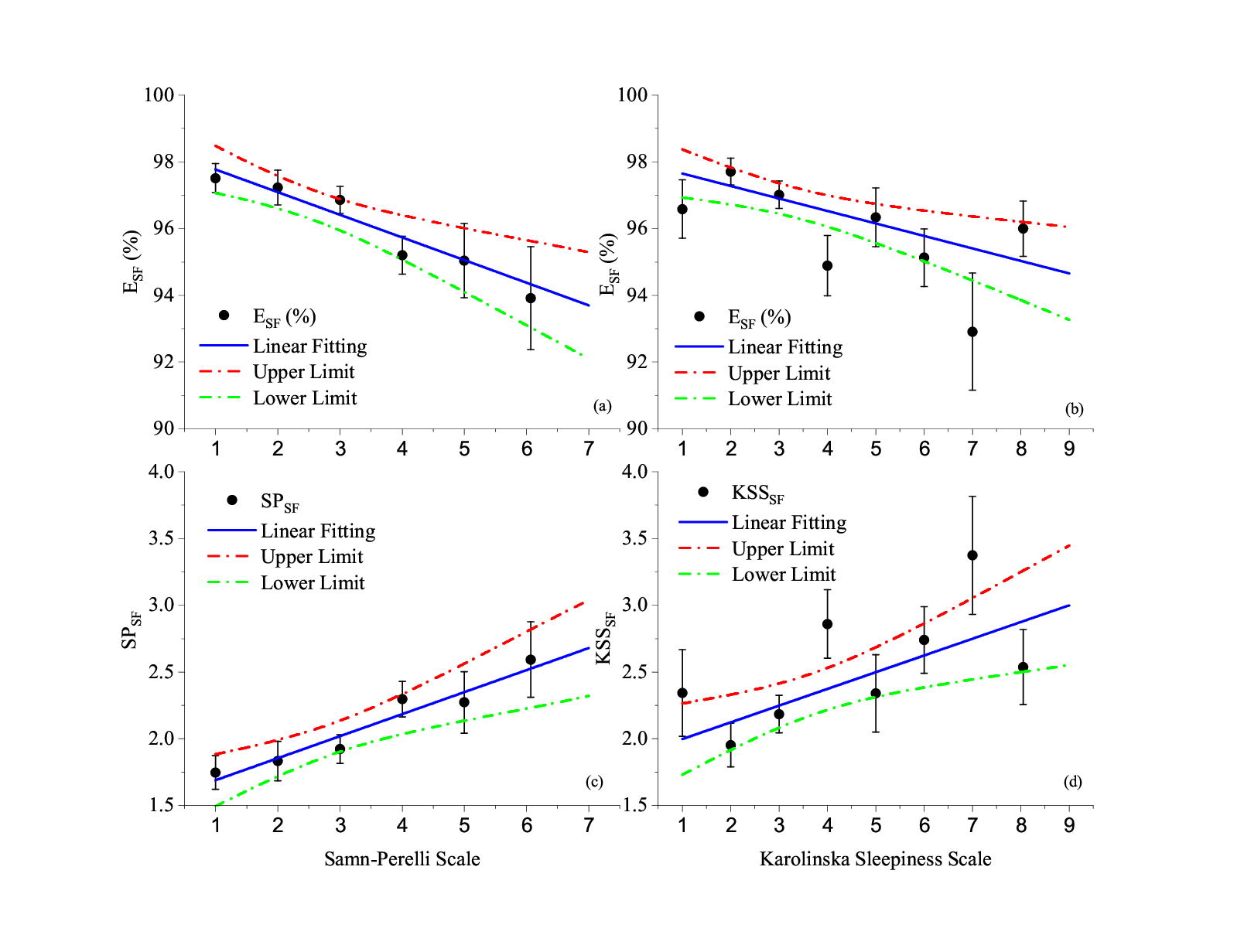}
\caption{\label{fig:Fig1}Upper panels: Self-rated SP (left) and KSS (right) scores versus average SAFTE-FAST Effectiveness (black circles) and associated standard errors (error bars). Lower panels: Self-rated SP (left) and KSS (right) scores versus average SAFTE-FAST SP (left) and KSS (right) (black circles) and associated standard errors (error bars). The blue lines represent the fitted functions and the dashed-dotted red and green lines represent, respectively, their upper and lower limits considering a 95\% confidence interval (CI) (N = 216, sample 3).}
\end{figure*}
\end{center}

\begin{center}
\begin{figure}
\includegraphics[scale=0.33]{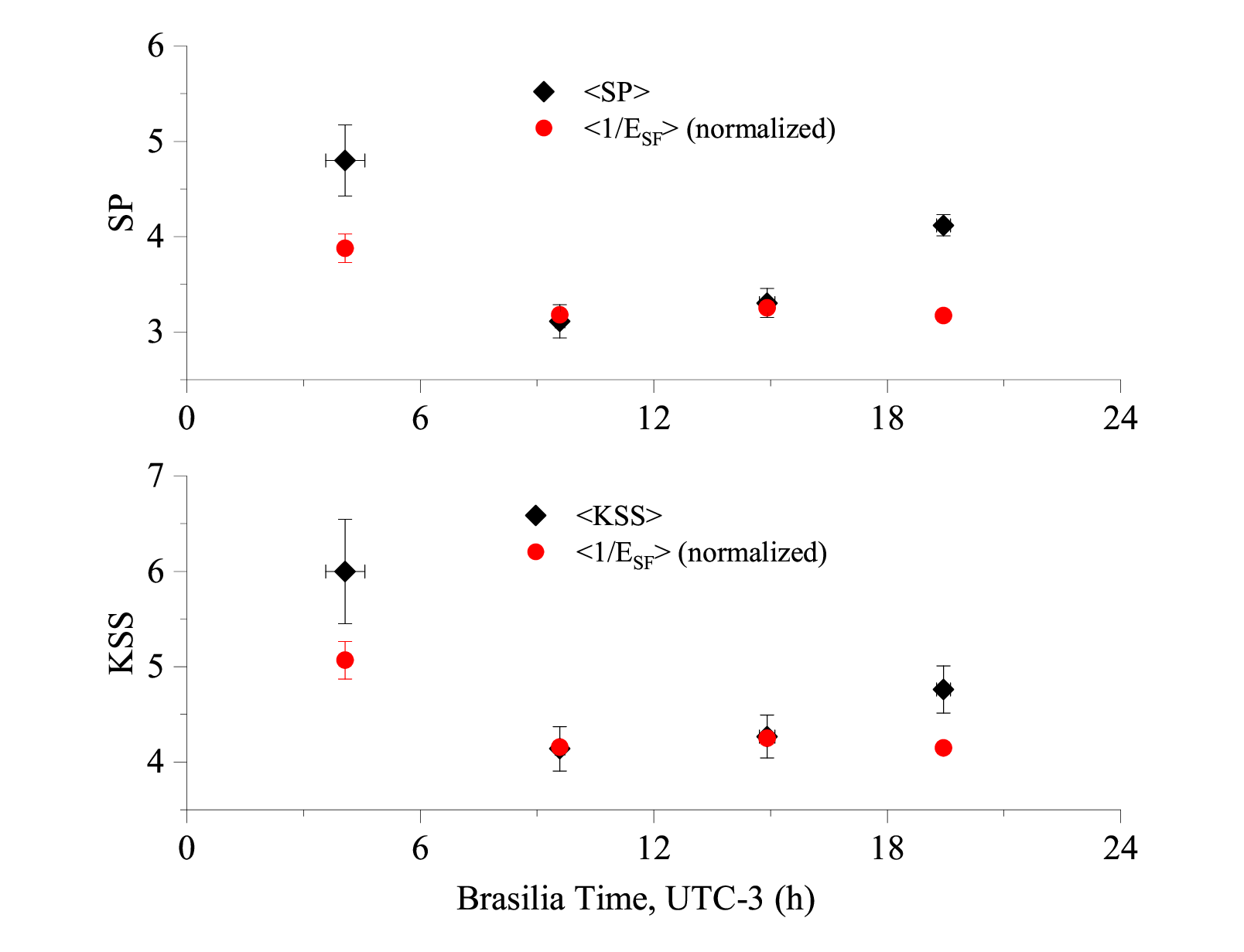}
\caption{\label{fig:Fig2} Averages of SP (upper panel) and KSS (lower panel) scores (black diamonds) and their standard errors (black error bars) as a function of the average response times for each clock-time group in Brasilia (Brazilian capital) time (UTC-3). Also shown the averages of $1/E_{SF}$, normalized to the SP (upper panel) and KSS (lower panel) averages (red circles) of groups 2 (morning) and 3 (afternoon), and their standard errors (red error bars) (N = 216, sample 3). Details in the text.}
\end{figure}
\end{center}

The SAFTE-FAST Effectiveness ($E_{SF}$) and Sleep Reservoir ($R_{SF}$) represent two important model outputs to estimate fatigue and sleep debt, respectively \citep{Rod2023}. They are usually calculated during critical phases of flight (departures and/or arrivals), indicating potentially hot spots of fatigue and/or excessive wakefulness during high demands of cognitive tasks.\\
The averages of these outputs - calculated for each 15-minute marker of sample 3 (N=216) - also demonstrate statistically significant Person's correlations with the self-rated fatigue and sleepiness scores from the questionnaire: SPx$E_{SF}$ ($\rho=-0.242, p<0.001$), SPx$R_{SF}$ ($\rho=-0.320, p<0.001$), KSSx$E_{SF}$ ($\rho=-0.184, p=0.007$) and KSSx$R_{SF}$ ($\rho=-0.168, p=0.014$). These findings show that the higher the SF Effectiveness (or Sleep Reservoir), the lower are the SP and KSS scores reported by the participants.\\
Linear relationships were also obtained using the LSM \citep{helene2016useful} and grouping the SP and KSS results into 6 and 8 intervals, respectively (sample 3, see Methods). Fig.\ref{fig:Fig1} presents the fittings (blue lines) of SP (upper-left panel) and KSS (upper-right panel) versus $E_{SF}$, as well as, SPx$\text{SP}_{SF}$ (lower-left panel) and KSSx$\text{KSS}_{SF}$ (lower-right panel). The dashed-dotted red and green lines represent, respectively, the upper and lower limits of the fitted function considering a 95\% Confidence Interval (CI). The results of the fittings are summarized in Table \ref{tab:table1}.\\
Circadian oscillations in the SAFTE-FAST Effectiveness and self-rated SP and KSS scores were also investigated by splitting sample 3 into four clock-time periods, herein defined as: dawn (from 0h00 to 05h59, group 1), morning (from 06h00 to 11h59, group 2), afternoon (from 12h00 to 17h59, group 3) and evening (from 18h00 to 24h59, group 4), all considering Brasilia time (UTC-3). This choice is particularly useful since night shifts are defined by Brazilian regulations \citep{Brasil2017, Anac2019} as any shift encompassing any fraction between 0h00 and 05h59, with the other groups defined with the same time interval. For each clock-time group, the averages of the response times and associated standard errors were calculated. Since the relative fatigue risk is proportional to the inverse of $E_{SF}$ \citep{Rod2023} - which derives from the finding that $1/E_{SF}$ also scales with the relative likelihood of human-error rail road accidents \citep{rod2020} - it is plausible also to compare this transformed variable with the self-rated SP and KSS scores. These results are presented in Fig.\ref{fig:Fig2} both for the SP (left) and KSS (right) scores, where the averages of $1/E_{SF}$ (with $E_{SF}$ in decimal units) have been multiplied by a normalization constant that fit the SP (and KSS) averages in groups 2 and 3 simultaneously. Such criterion was adopted after retaining the null hypothesis (Mann-Whitney tests for independent samples) between groups 2 (morning) and 3 (afternoon) both for SP ($p=0.310$) and KSS ($p=0.669$) and also considering the flat behaviour of the average values of $1/E_{SF}$ for these periods of the day.\\
Following the same steps described elsewhere \citep{Rod2023}, the SAFTE-FAST model was also used to calculate the average values of minimum effectiveness $EM_{C}$, minimum sleep reservoir $RM_{C}$, and fatigue hazard area $FHA_{C}$, all within critical phases of flight, considering a standardized 28-days epoch for all rosters of 2023 (sample 4). The results of this calculation are presented in Appendix A.

\begin{table} [h]
\caption{\label{tab:table1}
Slopes and intercepts of the fitted functions presented in Fig.\ref{fig:Fig1}. Also shown the number of degrees of freedom (N.D.F.), the $\chi^{2}$ and the probability ($p$) of exceeding the $\chi^{2}$.}
\begin{ruledtabular}
\begin{tabular}{cccccc}

\multirow{2}{*} {Function}   & \multicolumn{4}{c}{Fitting parameters}\\
& Slope &Intercept & N.D.F. & $\chi^{2}$ & $p$(\%)\\
 \hline 
SPx$E_{SF}$ & -0.68$\pm$0.18\%& 98.45$\pm$0.51\% & 4 & 2.57 & 63.1\\
KSSx$E_{SF}$ & -0.37$\pm$0.12\%&98.03$\pm$0.47\% & 6 & 10.06 & 12.2 \\
SPxSP$_{SF}$ & 0.165$\pm$0.043& 1.52$\pm$0.14 & 4 &1.94 & 74.6 \\
KSSxKSS$_{SF}$ & 0.125$\pm$0.041& 1.87$\pm$0.17& 6&10.02&12.4 \\

\end{tabular}
\end{ruledtabular}

\end{table}

\subsection{Pilot errors and sleepiness scores versus time of the day}

\begin{center}
\begin{figure}
\includegraphics[scale=0.33]{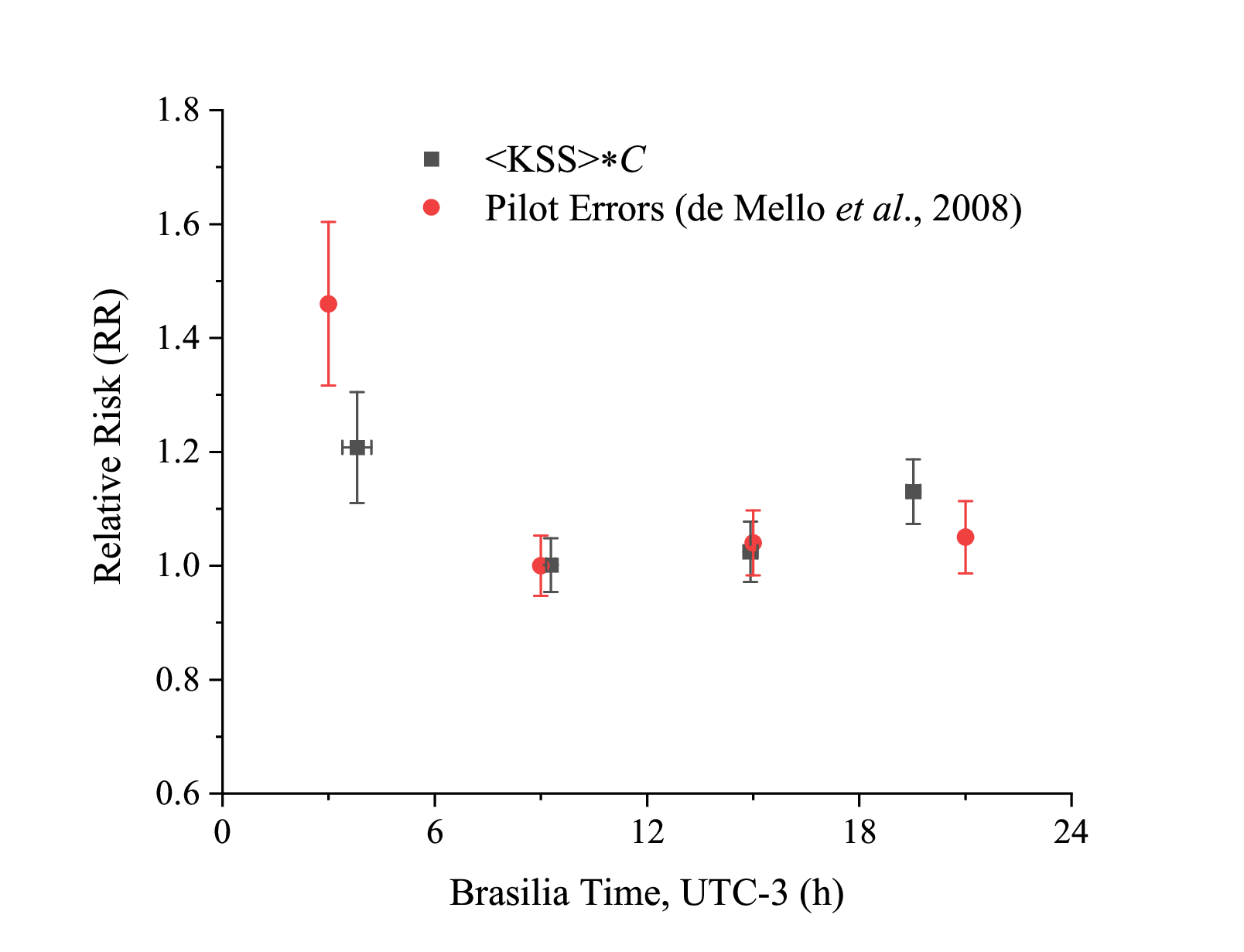}
\caption{\label{fig:Fig3} Relative Risk of pilot errors in the cockpit for the four clock-time groups (red circles with error bars) and the fitted averages of KSS (black squares with error bars) as a function of the average response times for each clock-time group, all in Brasilia (Brazilian capital) time (UTC-3) (N = 263, sample 2). Details in the text.}
\end{figure}
\end{center}

This section investigates the associations between the sleepiness scores from \textit{Fadigômetro} questionnaire (KSS) with pilot errors in the cockpit \citep{Mello2008}, as a function of the time of the day. The pilot errors were collected in a large Brazilian airline in 2005 via the analysis of 155,327 flight hours using the Flight Data Monitoring (FDM) system. Overall, de Mello et al. \citep{Mello2008} found 1065 pilots errors, distributed along the time of the day in the same time periods adopted in the previous section.\\
The KSS scores were collected within the same time periods, but including the full sample with questionnaire responses and rosters (sample 2, N = 263). The average values of KSS for each clock-time group were fitted to the relative risk of pilot errors (normalized to unit within 06h00 to 11h59) using the Least Square Method. The uncertainty in the average KSS was assumed as its standard errors, whereas the uncertainty in the relative risk of pilot error, herein denoted as relative risk ($RR$), was assumed as $\sigma _{RR} = RR\cdot\sqrt{N}/N$, where $N$ stands for the total number of errors (this assumption holds if the distribution of pilot errors along time follows a Poisson distribution). The fitted constant to the average KSS so obtained was $C=0.2389 \pm0.0098 $ with a $\chi^{2}$ of 3.03 and 3 degrees of freedom ($p = 0.386$). Fig.\ref{fig:Fig3} presents the fitted KSS scores (black squares) for the corresponding average values of the response times for each clock-time group and the relative risk of pilot errors (red circles) \citep{Mello2008} centered at each clock-time group.

\subsection{Workload patterns and self-rated fatigue and sleepiness scores}
\begin{center}
\begin{figure*}
\includegraphics[scale=0.65]{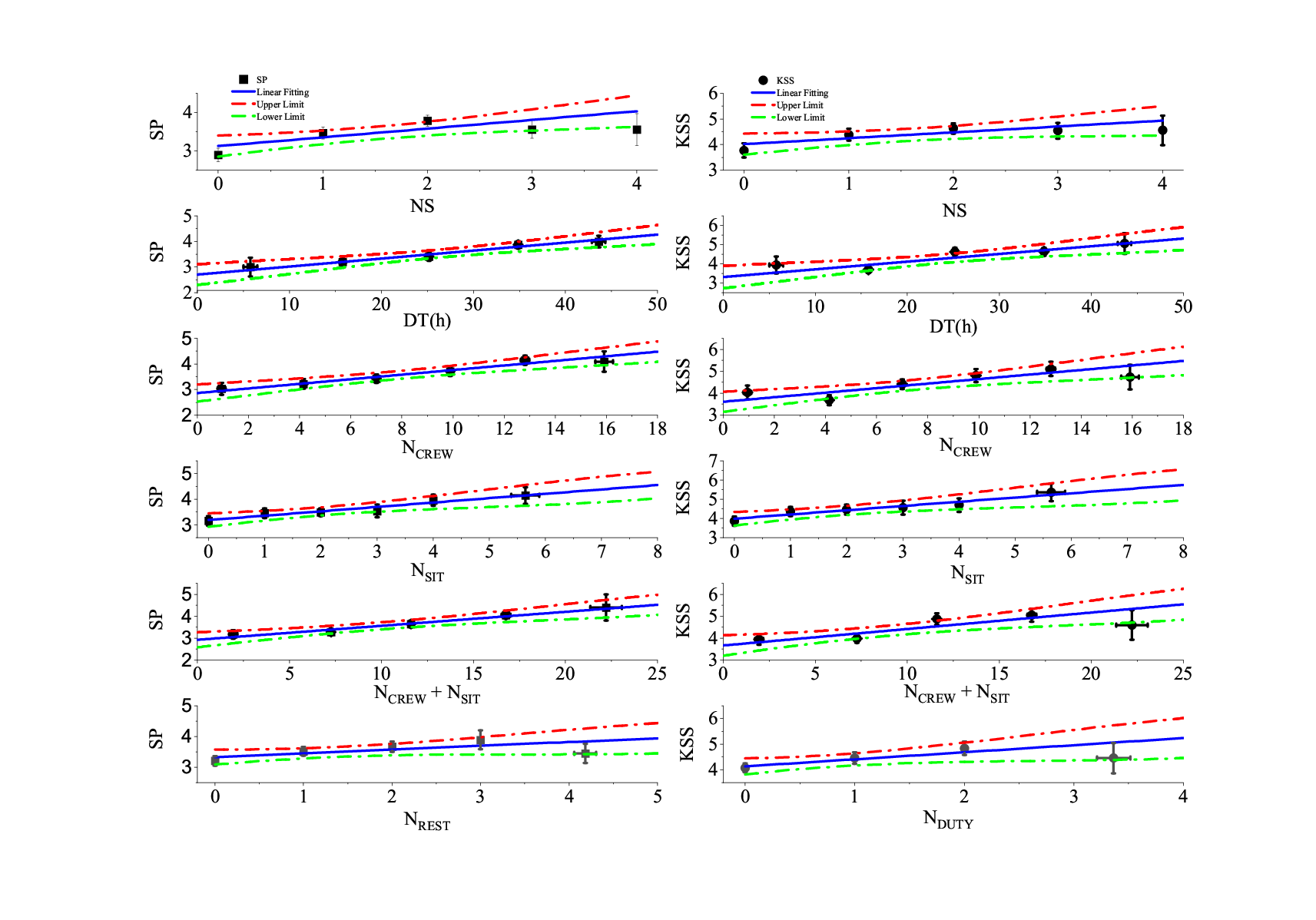}
\caption{\label{fig:Fig4} Average SP (left) and KSS (right) scores (data points) and their standard errors (error bars) as a function of workload metrics computed within 168 hours prior to the questionnaire response times (sample 2, N = 263). The fitted functions are shown by the solid blue lines, where the dashed-dotted red and green lines represent, respectively, their upper and lower limits considering a 95\% CI. Details in the text.}
\end{figure*}
\end{center}

To shed light on the complex interactions between workload patterns and fatigue/sleepiness outcomes, several rosters metrics of sample 2 (N=263) were carefully analysed through a dedicated computer algorithm within the 168 hour period prior to the questionnaire response times (see Methods). The Person's correlations between all the independent metrics and the fatigue/sleepiness outcomes are presented in Table \ref{tab:table2}. Among all metrics, the number of night-shifts ($N_{NS}$), total duty time [$DT(h)$], number of flight sectors ($N_{CREW}$) and number of sit times longer than one hour ($N_{SIT}$) have statistically significant correlations with both SP and KSS scores. Additionally, the number of short rests ($N_{REST}$) and long duties ($N_{DUTY}$) are also significantly correlated with SP and KSS, respectively.\\
Linear relationships were obtained between $N_{NS}$, $DT(h)$, $N_{CREW}$, and $N_{SIT}$, as well as the sum of flight sectors with long sit events $N_{CREW}+N_{SIT}$ with both SP and KSS scores, and also $N_{REST}$ versus SP and $N_{DUTY}$ versus KSS, as shown in Fig. \ref{fig:Fig4}. All the fittings were done by grouping the workload metrics into suitable intervals, where the number of bins ($N_{Group}$) equals the number of degrees of freedom ($N.D.F.$) plus two. The dispersions in the x-axis, whenever applicable, were also taken into account, such that the total variance of each data point (SP or KSS) could be written as:
\begin{equation}
\label{eqn:1}
\sigma_{y}^{2} = \langle\partial y/\partial x\rangle^{2}\cdot\sigma_{x}^{2}+\sigma_{y_{0}}^{2},
\end{equation}
where $\sigma_{y_{0}}$ is the standard error for SP (or KSS), $\sigma_{x}$ the standard error of the workload metric (x-error) and $ (\partial y/\partial x)$ the slope of the fitted function, which was calculated iteratively until its convergence. The results of all the fittings are shown in Table \ref{tab:table3}.

\begin{table}
\caption{\label{tab:table2}
Person's correlations ($\rho$) and respective $p-$values between the workload metrics and fatigue/sleepiness self-rated scores (SP and KSS) of sample 2 (N=263).}
\begin{ruledtabular}
\begin{tabular}{ccccc}
\multirow{3}{*} {Workload metric} & \multicolumn{4}{c}{Fatigue/sleepiness self-rated scores}\\
&  \multicolumn{2}{c}{SP} & \multicolumn{2}{c}{KSS}\\
& $\rho$ & $p-$value &  $\rho$ & $p-$value \\
\hline
$N_{NS}$	&\textbf{0.166}\footnotemark[1]&	\textbf{0.007}&	\textbf{0.125}\footnotemark[2]&	\textbf{0.042}\\
$N_{CNS}$	&0.056&	0.368&	0.044&	0.478\\
$N_{ES}$&	0.075&	0.223&	0.036&	0.561\\
$DT(h)$	& \textbf{0.208}\footnotemark[1]&	\textbf{0.001}&	\textbf{0.176}\footnotemark[1]&	\textbf{0.004}\\
$N_{CREW}$	&\textbf{0.260}\footnotemark[1]	&$<$\textbf{0.001}&	\textbf{0.211}\footnotemark[1]&	\textbf{0.001}\\
$N_{WOCL}$	&0.007&	0.907&	0.035&	0.572\\
$N_{SIT}$	&\textbf{0.180}\footnotemark[1]&	\textbf{0.003}&	\textbf{0.163}\footnotemark[1]&	\textbf{0.008}\\
$N_{REST}$	&\textbf{0.134}\footnotemark[2]&	\textbf{0.030}&	0.077&	0.215\\
$N_{DUTY}$	&0.096&	0.122&	\textbf{0.122}\footnotemark[2]&	\textbf{0.048}\\
$N_{CREW}+N_{SIT}$	&\textbf{0.253}\footnotemark[1]&	$<$\textbf{0.001}&	\textbf{0.209}\footnotemark[1]&	\textbf{0.001}\\

\end{tabular}
\end{ruledtabular}
\footnotetext[1]{The correlation is significant at the 0.01 level}
\footnotetext[2]{The correlation is significant at the 0.05 level}
\end{table}

\begin{table*}
\caption{\label{tab:table3}
Intercepts and slopes of the linear functions between workload metrics - computed within 168 hour prior to the questionnaire response times - and the self-rated fatigue (SP) and sleepiness (KSS) scores. Also shown the number of degrees of freedom (N.D.F.), the $\chi^{2}$ and the probability ($p$) of exceeding the $\chi^{2}$ for each fitting.}
\begin{ruledtabular}
\begin{tabular}{cccccc}

\multirow{2}{*}{Function} &  \multicolumn{5}{c}{Fitting parameters}\\
& Intercept & Slope & N.D.F. & $\chi^2$ & $p$ (\%)\\
\hline
$DT$xSP &2.71$\pm$0.22&	0.0312$\pm$0.0075 $h^{-1}$& 3 & 0.74 & 86.3\\
$DT$xKSS & 3.31$\pm$0.31&	0.040$\pm$0.012 $h^{-1}$& 3 & 3.99 & 26.3\\
$N_{NS}$xSP & 3.13$\pm$0.14&	0.227$\pm$0.076& 3 & 6.77 & 8.0 \\
$N_{NS}$xKSS & 4.02$\pm$0.21&	0.23$\pm$0.11& 3 & 2.41 & 49.2 \\
$N_{CREW}$xSP & 2.86$\pm$0.17&0.090$\pm$0.019& 4 & 1.16 & 88.4 \\
$N_{CREW}$xKSS & 3.61$\pm$0.24&	0.103$\pm$0.029& 4 & 5.01 & 28.6\\
$N_{SIT}$xSP & 3.18$\pm$0.13&	0.171$\pm$0.052&  4 & 0.97 & 91.4 \\
$N_{SIT}$xKSS & 3.98$\pm$0.18&	0.220$\pm$0.074& 4 & 1.08 & 89.7 \\
($N_{CREW}+N_{SIT}$)xSP & 2.93$\pm$0.17&	0.064$\pm$0.015& 3 & 1.09 & 80.0\\
($N_{CREW}+N_{SIT}$)xKSS & 3.67$\pm$0.23&	0.075$\pm$0.022& 3 & 4.88 & 18.1\\
$N_{REST}$xSP & 3.33$\pm$0.12&	0.122$\pm$0.070& 3 & 3.08 & 38.0\\
$N_{DUTY}$xKSS & 4.13$\pm$0.16&	0.28$\pm$0.13& 2 & 1.56 & 45.8\\

\end{tabular}
\end{ruledtabular}
\end{table*}

The fittings presented in Fig. \ref{fig:Fig4} represent meaningful results to understand the relationships between several workload metrics and the expected SP (or KSS) average values, considering an accumulated period of 168 hours before the assessment. On the other hand, these findings do not clarify a crucial question regarding fatigue risk management: What is the relative probability of having a high fatigue (or sleepiness) score considering different workload profiles? So, to address this question we conducted risk ratio calculations comparing groups of rosters of sample 2 with different workload levels, also assuming that scores of $\texttt{SP} \geq 6$ or $\texttt{KSS} \geq 7$ do represent high fatigue or sleepiness outcomes \cite{samn1982, aakerstedt1990, aakerstedt2014}.\\
According to Shapiro-Wilk Normality tests, almost all metrics - except $DT$ ($p = 0.325$) - are unlikely to be generated by a normal distribution ($p\leq0.002$). Consequently, the sample was divided into two Groups, namely: $G_{-}$, which included all the workload values below the average (or at or below the medians), and $G_{+}$, which included all values above the averages (or medians) for each distribution. The results of these analyses are presented in Table \ref{tab:table4}, where we found statistically significant risk ratios between $G_{+}$ and $G_{-}$ for $DT$, considering both SP [7.111 (95\% CI: 1.659-30.481)] and KSS [2.844 (95\% CI: 1.503-5.384)] scores, as well as, for $N_{CREW}$ [2.515 (95\% CI: 1.399-4.519)] and $N_{CREW}+N_{SIT}$ [2.450 (95\% CI: 1.380-4.349)] considering KSS scores. Moreover, Table \ref{tab:table4} also shows threshold values above which there are less than 2.5\% of events in each distribution, together with the actual fraction of events above these cutoff parameters. As will be discussed in detail later, these thresholds are promising metrics do address fatigue and sleepiness mitigation strategies in aircrew rosters.

 \begin{table*}
\caption{\label{tab:table4}
Risk Ratios $G_{+}/G_{-}$ for high fatigue ($\texttt{SP} \geq 6$) and sleepiness ($\texttt{KSS} \geq 7$) scores and their 95\% confidence intervals for all workload variables with statistically significant person's correlations with SP and/or KSS (see Table \ref{tab:table2}). Also shown the mean (or median), the threshold values and the actual fraction of excluded events for each distribution. Details in the text.}
\begin{ruledtabular}
\begin{tabular}{cccccc}

\multirow{2}{*}{Variable} & \multicolumn{2}{c}{Risk Ratio ($G_{+}/G_{-}$)\footnotemark[1]} & Mean or & Threshold & Fraction of\\
& $\texttt{SP} \geq 6$ & $\texttt{KSS} \geq 7$ & median& values &excluded events \\
\hline
$DT$&$\textbf{7.111 (1.659 - 30.481)}$	&	\textbf{2.844 (1.503 - 5.384)}	&$25.4h$	& $43.9h$\footnotemark[2] & 2.3\%\\
$N_{NS}$&0.960 (0.287 - 3.209)	&	0.995 (0.495 - 2.003)	&	2&$ 4$\footnotemark[3] & 0\%\\
$N_{CREW}$&1.677 (0.658 - 4.270)	&	\textbf{2.515 (1.399 - 4.519)}	&	7&	$15$ & 1.9\%\\
$N_{SIT}$&1.691 (0.668 - 4.280)	&	1.249 (0.711 - 2.196)	&	2&$5$ & 2.3\%\\
$N_{CREW}+N_{SIT}$&1.810 (0.695 - 5.120)	&	\textbf{2.450 (1.380 - 4.349)}	&	9&	$19$ & 1.9\%\\
$N_{REST}$&1.521 (0.607 - 3.813)	&	0.718 (0.395 - 1.304)	&	1 &$4$& 0.8\%\\
$N_{DUTY}$&0.900 (0.304 - 2.666)	&	1.513 (0.866 - 2.644)	&	1 &$3$ & 1.5\%\\

 \end{tabular}
\end{ruledtabular}
\footnotetext[1]{The risk ratios and their 95\% confidence intervals were calculated using SPSS version 25.}
\footnotetext[2]{For practical purpose the $DT$ threshold was fixed at $44h$.}
\footnotetext[3]{The number of night shifts within 168 hours is limited by regulations to 4. Fixing its threshold to 3 yields a fraction of excluded events higher than 2.5\%.}
\end{table*}

\subsection{Estimating maximum cumulative fatigue and sleepiness scores from roster workload patterns: a novel Fatigue Risk Management approach}

In this section, a novel fatigue risk management approach is proposed in order to estimate the maximum cumulative fatigue and sleepiness scores by analysing roster workload patterns. The method also addresses the impacts of mitigation strategies on both operational safety (i.e: reduction in the predictive scores of SP and KSS), as well as on aircrew productivity (i.e: impacts on roster production). The approach is based on the implementation of mitigations in some key workload metrics, such as $DT$, $N_{CREW}$ and $N_{CREW}+N_{SIT}$, keeping them below or equal the threshold values that exclude less than 2.5\% of the higher figures of sample 2 (see Table \ref{tab:table4}). These workload metrics were chosen given their statistically significant risk ratios for either high fatigue ($\texttt{SP} \geq 6$) or sleepiness ($\texttt{KSS} \geq 7$) scores found between $G_{+}$ and $G_{-}$. Considering that these thresholds impact in less than 2.5\% of each distribution one would expect small impacts in aircrew productivity during rostering optimization processes.\\
The analyses are then performed via the implementation of the following steps:  
\begin{enumerate}
\item First, we pick all non-augmented crew rosters of 2023 (sample 4) using a standardized 28-days epoch for all months;
\item Second, we calculate the expected values of SP and KSS for each 168-hour loop and each workload metric using the corresponding fitted functions (see Table \ref{tab:table3}) plus a random value (negative or positive) normally distributed around zero ($\mu = 0$) with standard deviation equals to $\sigma_{f}$, which is the propagated uncertainty of the fitted functions ($\texttt{95\% CI} \approx 1.96\sigma_{f}$);
\item Third, we collect the maximum value of SP and KSS considering all 168-hour loops within 28 days for each roster of 2023 (total of 3,610,145 loops);
\item Fourth, we perform the same calculations of the previous step, but include a mitigation strategy that restricts some key workload metrics at their thresholds shown in Table \ref{tab:table4}, such that $DT\leq 44 h$, $N_{CREW}\leq 15$ and $N_{CREW}+N_{SIT} \leq 19$ for each 168-hour loop. 
\end{enumerate}

\begin{center}
\begin{figure*}
\includegraphics[scale=0.65]{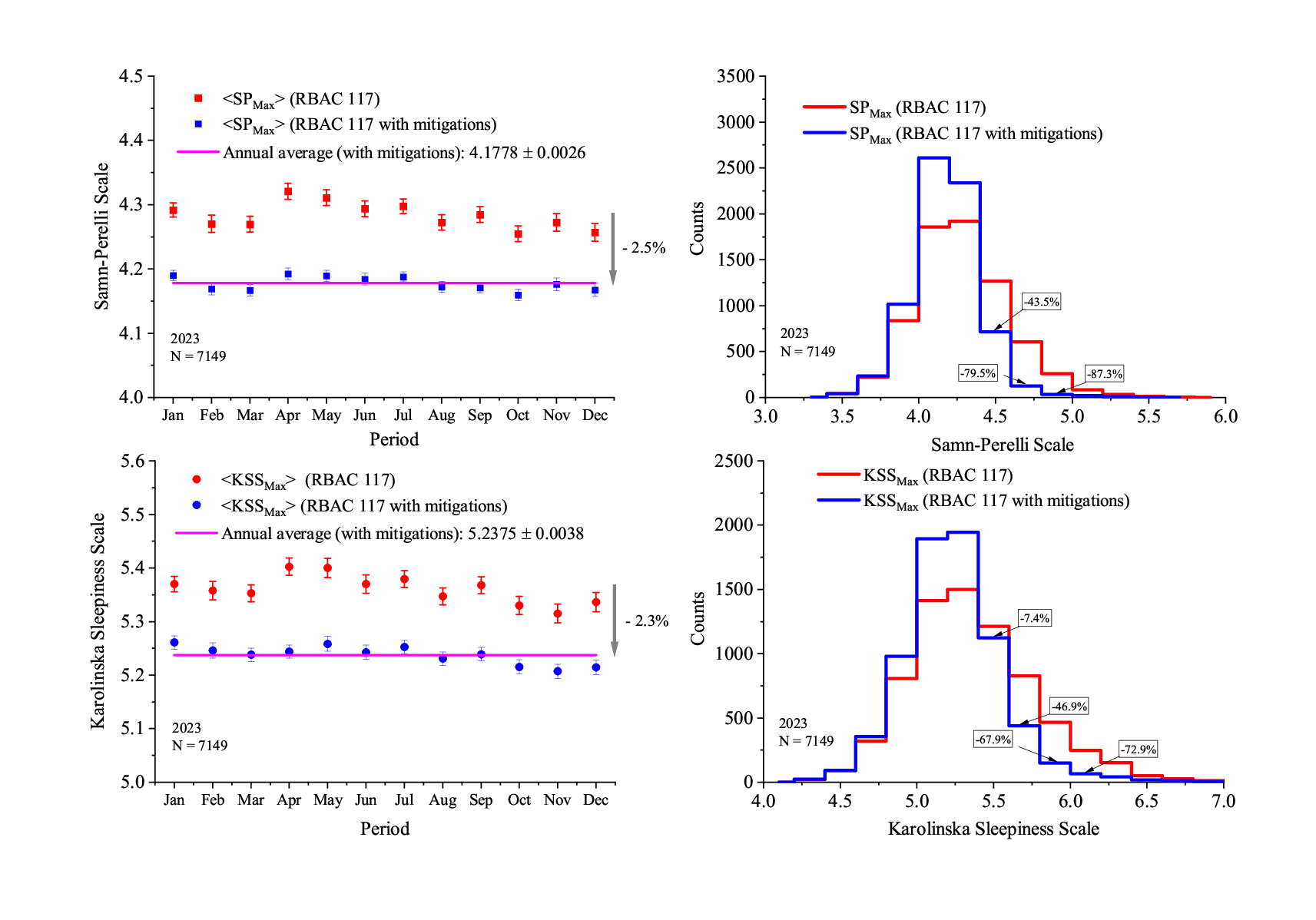}
\caption{\label{fig:Fig5} Left upper/lower panels: Average maximum SP/KSS scores and their standard errors under the current RBAC 117 rules (red squares/circles) and with the proposed mitigations (blue squares/circles). Also shown the 2023 annual averages with mitigations (magenta lines) both for SP (upper) and KSS (lower) scores. Right upper/lower panels: Maximum SP/KSS distributions under the RBAC rules (red histograms) and with the proposed mitigations (blue histograms).}
\end{figure*}
\end{center}

Under this framework, it is possible to estimate the maximum values of SP and KSS (with or without mitigations) by the combination of the analysis of all workload variables in each successive 168-hour loops for all the rosters of 2023 (N = 7149) and the fitted functions with their respective standard errors found in the previous section (see Fig.\ref{fig:Fig3} and Table \ref{tab:table3} for details). \\
Considering the current RBAC 117 rules, the predicted average values and standard errors of the maximum SP/KSS scores for each 28-day epochs of 2023 are presented by the red squares/circles in the upper/lower left-hand panels of Fig.\ref{fig:Fig5}. The blue squares/circles in the upper/lower left-hand panels represent the average SP/KSS maximum values for the same epochs after applying the mitigations for $DT$, $N_{CREW}$ and $N_{CREW}+N_{SIT}$. The magenta lines show the mitigated annual averages for $\texttt{SP}_{MAX}$ $(4.1788\pm0.0026)$ and $\texttt{KSS}_{MAX}$ $(5.2375\pm0.0038)$, obtained via the Least Squares Method (SP: $\chi^{2}=17.4, N.D.F. = 11, p = 9.8\%$ and KSS: $\chi^{2}=19.5, N.D.F. = 11, p = 5.3\%$).\\
The distributions of the maximum SP/KSS scores considering all 7149 rosters of 2023 under the RBAC 117 rules are presented by the red histograms in the upper/lower right-hand panels of Fig.\ref{fig:Fig4}, whereas the blue histograms show the SP/KSS maximum scores distributions after applying the mitigations.\\
Considering all rosters of 2023 under RBAC 117 rules (without mitigation), the workload metrics of $DT$, $N_{CREW}$ and $N_{CREW}+N_{SIT}$ were associated with 70.1\%(60.6\%) of the highest predicted scores of SP(KSS), reinforcing the relevance of these variables in the proposed mitigation. An overall analysis of the workload metrics associated with the predicted highest values of SP and KSS can be found in Appendix B.\\     
Potential impacts of the proposed mitigations on aircrew productivity were also investigated, where we found that over the 3,610,145 loops of all rosters of 2023, only 3.9\%, 3.1\% and 3.8\% exceed the thresholds $DT\leq 44 h$, $N_{CREW}\leq 15$ and $N_{CREW}+N_{SIT} \leq 19$, respectively. Combining the three workload metrics, only 6.7\% of all loops exceed at least one of the proposed limitations. The complete analysis of this impact can be found in Appendix B.\\
For the sake of completeness, the workload metrics were also calculated considering a 28-days epochs for all rosters of 2023 (sample 4). The results are presented in Appendix C.
     
\section{Discussion} 

The self-rated fatigue (SP) and sleepiness (KSS) scores obtained from the questionnaire responses showed statistically significant Person's correlations with SAFTE-FAST Effectiveness, $E_{SF}$ ($\rho=-0.242,p<0.001$ and $\rho=-0.184,p=0.007$, respectively). On the other hand, the linear relationships so obtained (upper panels of Fig.\ref{fig:Fig1}) demonstrate small variations of $E_{SF}$ for large variations in SP (or KSS). In fact, varying the SP scores from 1 to 7 (or KSS from 1 to 9) yields a small decrease in $E_{SF}$, typically from $\sim$ 97.8\% to 93.7\% (or $\sim$ 97.7\% to 94.7\%). Furthermore, the slopes found between SP and $\texttt{SP}_{SF} (0.165\pm0.043)$ and KSS and $\texttt{KSS}_{SF}(0.125\pm0.041)$ (see Table \ref{tab:table1}) make clear that just small fractions of the variations in the self-rated scores are being reflected in the model outcomes (ideally, one would expect slopes close to unit). Such results demonstrate the limitations of the model to accurately reproduce perceptions of fatigue and sleepiness, as well as, strong evidences of model outcomes with high false negative for fatigue/sleepiness. As a consequence, the model results should be used with caution and not as the sole source of information to rule out the likelihood of potentially high levels of fatigue and/or sleepiness in actual aircrew rosters. A possible explanation for this finding would be that the model does not take workload effects into account when calculating $E_{SF}$ (or $\texttt{SP}_{SF}$ and $\texttt{KSS}_{SF}$), probably causing an underestimation of fatigue and/or sleepiness perceptions. Moreover, the inverse of $E_{SF}$ in sample 3 does not vary considerably between 06h00 and 24h00, in contrast with the circadian oscillations found for SP and at some extent also for KSS.(see Fig. \ref{fig:Fig2}). In fact, the averages of $1/E_{SF}$, normalized to the SP (or KSS) scores of groups 2 (from 06h00 to 11h59) and 3 (from 12h00 to 17h59) present modest peaks within 0h00 and 05h59, but do not reproduce the increase in fatigue (or sleepiness) perceptions between 18h00 and 23h59. Once again, the peak in the self-rated SP - and qualitatively also in KSS - during the evening could be associated with the effects of workload, a feature not included in the calculation of $E_{SF}$.\\
The circadian oscillations found in the self-rated sleepiness scores for sample 2 (N = 263) are quantitatively associated with the relative risk of pilot errors in the cockpit (see Fig. \ref{fig:Fig3}), showing the relevance of a subjective sleepiness score to reproduce objective FDM data ($\chi^2 = 3.03, N.D.F. = 3 \texttt{ and }p = 0.386$). 
Despite the large uncertainties (error bars) in pilot errors and KSS scores (see Limitations), this finding suggests that errors in the complex cockpit environment are associated with perceptions of sleepiness.\\
The linear relationships found between seven workload metrics - computed within 168 hours prior to the questionnaire responses - and both SP and KSS scores of sample 2 (see Fig. \ref{fig:Fig4} and Table \ref{tab:table3}) demonstrate how fatigue/sleepiness perceptions increase with respect to roster's characteristics. Furthermore, the risk ratios for high fatigue ($\texttt{SP} \geq 6$) or sleepiness ($\texttt{KSS} \geq 7$) scores when comparing groups $G_{+}$ and $G_{-}$ (see Table \ref{tab:table4}) indicate statistically significant figures for the total duty time ($DT$) both for SP and KSS, and for the number of flight sectors ($N_{CREW}$) and the  aggregated  metric ($N_{CREW}+N_{SIT}$) with KSS. Such finding makes clear the benefits of mitigation strategies to eliminate extreme situations via the inclusion of threshold values that exclude only few percent (herein fixed at or below 2.5\%) of the respective metric distributions (see Table \ref{tab:table4}). Obviously, these thresholds based on a cut-off of less than 2.5\% of each distribution are arbitrary and likely extreme values, but they demonstrate the potential of this new approach to mitigate the risk of high scores of fatigue and sleepiness with the likelihood of minimal impact on roster optimization processes. In this sense, for the purpose of fatigue risk management, operators should analyse the benefits of the mitigations, increasing cut-off values and verifying their impacts on aircrew productivity (see the Safety Recommendations section).\\
The analysis of all 2023 rosters (sample 4, N = 7149) via the linear relationships found for sample 2 (N = 263) propitiated a robust method to estimate maximum values of fatigue (SP) and sleepiness (KSS), as presented in Fig.\ref{fig:Fig5}. The novel approach calculates the maximum fatigue and sleepiness scores for each 168-hour loop and selects the global maximum within a 28-day epoch for each roster. It also takes into account the uncertainties of the fitted functions (see Fig. \ref{fig:Fig4}) using a Monte Carlo sampling technique, thus ensuring that the final results reflect the covariance matrix of the fitted parameters and, ultimately, the variance of the fitted data of sample 2 (see also the Limitations section).\\
The results of the averaged maximum values of SP (KSS) for each month of 2023 are presented by the red squares (circles) in the left-upper (lower) panels of Fig. \ref{fig:Fig5}. The predicted averages of $\texttt{SP}_{\texttt{MAX}}$ ($\texttt{KSS}_{\texttt{MAX}}$) scores after mitigations are 2.5\% (2.3\%) lower, when compared with the actual RBAC 117 rules, as shown by the fitted magenta lines in the left-upper (lower) panels of Fig.\ref{fig:Fig5}. Indeed, the mitigated annual averages found for $\texttt{SP}_{\texttt{MAX}}$ ($4.1778 \pm 0.0026$) and $\texttt{KSS}_{\texttt{MAX}}$ ($5.2375 \pm 0.0038$) do represent very precise figures, with uncertainties of 0.06 and 0.07\%, respectively. Despite the modest decrease in the annual averages of $\texttt{SP}_{\texttt{MAX}}$ ($\texttt{KSS}_{\texttt{MAX}}$) after applying the mitigations, their effectiveness and benefits to reduce higher scores are evident via the inspection of the upper (lower) right panels of Fig. \ref{fig:Fig5}. The blue histograms show a significant reduction of the higher values of the distributions both for $\texttt{SP}_{\texttt{MAX}}$ (upper) and $\texttt{KSS}_{\texttt{MAX}}$ (lower), when compared with the non-mitigated RBAC 117 regulations (upper and lower red histograms, respectively). The combined thresholds for $DT$, $N_{CREW}$ and $N_{CREW}+N_{SIT}$ are exceed in only 6.7\% of the total 3,610,145 loops analysed, thus demonstrating the applicability of the mitigations with minimal impact on rosters production.\\

\section{Limitations of the study}

Despite that this study includes several rosters from the same individual, the self-rated SP and KSS assessments were made just once in time. Consequently, the individual variations along time and with respect to different roster's characteristics could not be captured. Such cross-sectional method limits the statistical robustness of the sample when associating workload metrics with fatigue and sleepiness outcomes (sample 2). In fact, considering the multinomial character of both SP and KSS scores, one would expect Sigmoid shapes between these data and the workload metrics, instead of linear relationships, which are not bounded within asymptotic maximum/minimum values. On the other hand, the limitations related with the linear assumptions were at some extent taken into account in the uncertainty ranges of the fitted functions, which depend on the variance matrix of the data. In this regard, it is still feasible and very insightful to extrapolate the results found for sample 2 for the analysis of all 7149 rosters of 2023. Ideally, longitudinal experiments with objective sleep measurements and multiple fatigue and sleepiness assessments for the same individual - as the one proposed recently \cite{sampaio2023} - are very welcome to capture individual random effects and provide a better understanding of the complex relationships between workload and fatigue or sleepiness.\\
The association of pilot errors in the cockpit with self-rated KSS scores (see Fig.\ref{fig:Fig3}) should be interpreted with caution. Uncertainties in the data (error bars) - especially during the dawn ($\sim10\%$) - hinder a more rigorous test. Despite being statistically consistent, the relative risk of pilot error increases approximately 46\% between 00:00 and 05:59 compared to baseline results found between 06:00 and 11:59, but KSS scores increase approximately 20\%. In this sense, a different experimental design, with multiple responses from the same individual, focused only on shift days, would provide more robust statistics, with the possibility of having more granular clock-time intervals.\\
Albeit being a useful approach to estimate cumulative self-rated SP and KSS scores based on roster's workload metrics, the results herein obtained might have been affected by the COVID-19 outbreak, mostly in 2021 and 2022. In addition, the proposed task-related workload metrics do not exhaust the cognitive demands in the complex aviation environment that also depends on weather conditions, air traffic, abnormal airplane configurations, flight disruptions, to name a few.\\
Other minor limitation of the study is related with the non-inclusion of home standby duties, which, as mentioned previously \cite{Rod2023}, are indistinguishable from days off in our analyses.
  
\section{Safety recommendations}
This section summarizes our main findings and the associated safety recommendations applicable for non-augmented minimum crew passenger flights dedicated to short/medium haul operations in Brazil. We also provide insights for best practices for the use of bio-mathematical models, as well as a novel data-driven fatigue risk management approach.

\subsection{Mitigations in the RBAC 117 rules}
Considering our predictions for cumulative fatigue and sleepiness, it is recommended that the following limitations are considered in planned and executed rosters:
\begin{enumerate}
\item Total Duty Time ($DT$): The total duty time, excluding home standby duties, should be limited to 44 hours for each 168-hour period in the roster. This limit should not apply for long haul international flights; 
\item Flight Sectors ($N_{CREW}$): The number of flight sectors should be less than 16 for each 168-hour period of the roster;
\item Flight Sectors plus Sit Times longer than one hour ($N_{CREW}+N_{SIT}$): The number of flight sectors plus the number of sit times longer than one hour should be less than 20 for each 168-hour period in the roster. 
\end{enumerate}

\subsection{Best practices of using bio-mathematical models}
According to ICAO DOC 9966 \citep{ICAO2016}, bio-mathematical models ``do not constitute an FRMS on their own, but are only one tool of many that may be used within an FRMS". Additionally, ICAO brings an implication for States, which ``should not rely on bio-mathematical models as the sole means of evaluating the effectiveness of a Service Provider’s FRMS". In this regard, bio-mathematical models should not be used as the sole information for ``GO" decisions in the following scenarios (given their high false negative for fatigue):
\begin{enumerate}
\item Decisions that allow more flexible prescriptive limits;
\item Decisions that disqualify a fatigue report from aircrew;
\item Decisions that discard the likelihood of fatigue in rosters with relevant workload.
\end{enumerate}

\subsection{Data-driven Fatigue Risk Management Approach}
For the purpose of managing workload, fatigue and sleepiness, Operators and Scheduling Personnel should:
\begin{enumerate}
\item consider the workload metrics of $DT$, $N_{CREW}$ and $N_{CREW}+N_{SIT}$ as Key Performance Indicators (KPI) during the rostering optimization processes. These parameters should be kept as low as reasonably achievable for every 168-hour period for each individual;
\item consider limiting the total duty time to 44 hours or less, the total number of flight sectors to less than 16 and the sum of flight sectors and sit periods longer than one hour to less than 20 at each 168-hour loop during the rostering optimization processes;
\item evaluate the benefits of fatigue risk mitigations in $DT$, $N_{CREW}$ and $N_{CREW}+N_{SIT}$ - computed at every 168-hour period - via the analysis of their relationships with the available operational indicators, such as, exceedances in Flight Data Monitoring, hard landings, missed approaches, as well as medical records, among others.
\end{enumerate}
   
\section{Conclusions}
This work proposes a novel approach to investigate the complex relationships between workload metrics and self-rated fatigue (Samn-Perelli, SP) and sleepiness (Karolinska, KSS) scores. The study also addresses the limitations of bio-mathematical models that may provide high false negative results that do not accurately reflect the perceptions of fatigue and sleepiness, which are most likely associated with roster's characteristics, in addition to the main model ingredients related with the homeostatic sleep process and circadian rhythms.\\
On the other hand, the KSS averages as a function of the time of the day actually fit objective measurements of the relative risk of pilot errors in the cockpit \citep{Mello2008}, thus providing a reliable subjective indicator for risk assessment.\\
The linear relationships found between workload metrics and SP (or KSS) scores provided a consistent method to estimate peak values of accumulated fatigue and sleepiness for every 168-hour loop. The risk ratios for high fatigue ($\texttt{SP} \geq 6$) or sleepiness ($\texttt{KSS} \geq 7$) comparing groups $G_{+}$ and $G_{-}$ (see Table \ref{tab:table4}) are statistically significant for the total duty time ($DT$) both for SP and KSS, and for the number of flight sectors ($N_{CREW}$) and the sum of flight sectors with sit times longer than one hour ($N_{CREW}+N_{SIT}$) with KSS. These workload variables are also associated with 70.1\% (60.6\%) of the highest predicted scores of SP (KSS) considering all rosters of 2023, representing key metrics for safety recommendations either for the regulatory framework, as well as for fatigue risk management approaches. Applying the mitigations $DT\leq 44 h$, $N_{CREW}\leq 15$ and $N_{CREW}+N_{SIT} \leq 19$ for every 168-hour interval yields a significant decrease in the higher values of SP (and KSS) with minimal impact (6.7\% of all loops of 2023) on aircrew productivity.\\ 
Further longitudinal experiments with objective sleep measurements and multiple fatigue and sleepiness assessments are very welcome to provide more robust results with the possibility of implementing daily analyses and mitigations, in addition to the cumulative (weekly) approach proposed here.

\section{Credit author statement}
\textbf{Tulio E. Rodrigues:} Conceptualization, Methodology, Validation, Formal analysis, Data Curation, Writing - Original Draft and Project Administration. \textbf{Eduardo Furlan}: Validation, Data Curation, Writing - Review \& Editing. \textbf{André F. Helene}: Conceptualization, Methodology, Writing - Review \& Editing, Supervision. \textbf{Otaviano Helene}: Methodology, Formal analysis, Writing - Review \& Editing, Supervision. \textbf{Eduardo Pessini}: Data Curation, Writing - Review \& Editing. \textbf{Alexandre Simões}: Writing - Review \& Editing. \textbf{Maurício Pontes}: Writing - Review \& Editing.  \textbf{Frida M. Fischer}: Conceptualization, Methodology, Writing - Review \& Editing, Supervision.

\section{Acknowledgment}

We would like to thank Dr. Izabela Sampaio for the fruitful scientific discussions and for the kind review of this manuscript, Mr. Denys Sene, from IASERA, for the development and support to the \textit{Fadigometro} web-based platform and Mr. Marcelo Brugnera for the development of the \textit{Fadigometro} App. We also thank the National Commission of Human Fatigue (CNFH), the Aeronautical Accidents Investigation and Prevention Center (CENIPA) and the National Committee for the Prevention of Aeronautical Accidents (CNPAA) for their support and endorsement of the study within the aviation community and all the anonymous crew members who voluntarily participated in the study.

\section{Funding}  
This project is equally funded by the Brazilian Association of Civil Aviation Pilots (ABRAPAC), Gol Aircrew Association (ASAGOL) and LATAM Aircrew Association (ATL).

\appendix

\section{SAFTE-FAST results for all rosters of 2023 (sample 4)}

In this section we present the SAFTE-FAST results for the average values and respective standard errors for the minimum effectiveness $EM_{C}$ (\%), minimum sleep reservoir $RM_{C}$ (\%), and fatigue hazard area $FHA_{C}$ (min), all computed within critical phases of flight, which include the first and last 30 minutes of each flight sector. The SAFTE-FAST runs were performed with software version 6.6 assuming a 60-minute commuting time from home to station, hotel to station and rest facility to station and vice-versa and the standard SAFTE-FAST Auto-Sleep controls. The details for the SAFTE-FAST parameters and criteria can be found elsewhere \citep{Rod2023}. Differently from our previous work that adopted 30-days epochs, the simulations performed here consider a standardized 28-days epoch for each month of 2023, which included duty periods starting after each first day at 0h00 and before the $29^{th}$ day at 0h00, all in Brasilia time (UTC-3 hours); except for February 2023, where the ending point was chosen as March 1st, 2023 at 0h00 Brasilia Time. The results for all months of 2023 (N = 7149, sample 4) are presented in Table \ref{tab:table5}.

\begin{table}
\caption{\label{tab:table5}
Averages and standard errors (SE) for minimum effectiveness $EM_{C}$ (\%), minimum sleep reservoir $RM_{C}$ (\%) and fatigue hazard area $FHA_{C}$ (min), calculated within critical phases of flight and 28-days epochs for all rosters of 2023 (sample 4) via the SAFTE-FAST model. Details in the text.}
\begin{ruledtabular}
\begin{tabular}{cccccc}
Period & N & Parameter & $EM_{C}$ (\%) & $RM_{C}$ (\%) &$FHA_{C}$ (min) \\
 \hline

\multirow{2}{*}{Jan-23} & \multirow{2}{*}{650} & Average &73.55	&78.17	&4.57\\
& & SE &0.20	&0.12	&0.21\\
\multirow{2}{*}{Feb-23} & \multirow{2}{*}{508} & Average &75.07	&78.95	&3.28\\
& & SE &0.23	&0.12	&0.21\\
\multirow{2}{*}{Mar-23} & \multirow{2}{*}{628} & Average &75.93	&79.11	&3.39\\
& & SE &0.27	&0.13	&0.21\\
\multirow{2}{*}{Apr-23} & \multirow{2}{*}{628} & Average &76.11	&79.22	&2.98\\
& & SE &0.26	&0.12	&0.19\\
\multirow{2}{*}{May-23} & \multirow{2}{*}{610} & Average &76.43	&79.42	&2.58\\
& & SE &0.25	&0.12	&0.19\\
\multirow{2}{*}{Jun-23} & \multirow{2}{*}{614} & Average &76.18	&79.25	&2.78\\
& & SE &0.25	&0.12	&0.19\\
\multirow{2}{*}{Jul-23} & \multirow{2}{*}{622} & Average &75.04	&78.73	&4.30\\
& & SE &0.27	&0.14	&0.28\\
\multirow{2}{*}{Aug-23} & \multirow{2}{*}{658} & Average &75.97	&79.23	&2.59\\
& & SE &0.24	&0.12	&0.16\\
\multirow{2}{*}{Sep-23} & \multirow{2}{*}{641} & Average &76.20	&79.39	&2.50\\
& & SE &0.23	&0.11	&0.16\\
\multirow{2}{*}{Oct-23} & \multirow{2}{*}{580} & Average &75.69	&79.21	&3.01\\
& & SE &0.27	&0.14	&0.19\\
\multirow{2}{*}{Nov-23} & \multirow{2}{*}{547} & Average &76.17	&79.70	&2.60\\
& & SE &0.25	&0.11	&0.16\\
\multirow{2}{*}{Dec-23} & \multirow{2}{*}{464} & Average &74.95	&79.05	&3.19\\
& & SE &0.26	&0.13	&0.19\\

 \end{tabular}
\end{ruledtabular}
\end{table}

\section{Workload factors associated with the highest predicted values of SP and KSS for all rosters of 2023 (sample 4)}

Considering all 7149 rosters of 2023, the total duty time, the number of flight sectors and the sum of flight sectors with sit periods longer than one hour were responsible for 25.7(25.6 \%), 26.1(20.2 \%) and 18.3(14.8 \%) of the highest SP(KSS) scores, respectively. The number of night shifts and duties above 9 hours were associated with 15.6(12.3 \%) and 10.5(15.3 \%) of the predicted peak values of SP(KSS), respectively. The number of sit times longer than one hour and the number of rest periods shorter than 16 hours were associated with 11.9 \% and 3.8\% of the highest KSS and SP scores, respectively. Table \ref{tab:table6} brings the percentages of rosters associated with the workload metrics that produced the highest predicted scores of SP and KSS within 28-days epochs for all rosters of 2023.\\
In order to estimate the impact of the proposed mitigations for the rostering optimization processes we have calculated the fraction of all 168-hour loops that exceed the thresholds $DT>44$ h, $N_{CREW}>15$ and $N_{CREW}+N_{SIT}>19$. The results are presented in Table \ref{tab:table7} for all months of 2023 and show that only a minor fraction of all loops overshoots the proposed mitigations. Combining the mitigations we found an overall impact of 6.7\% considering all 3,610,145 loops of 2023 (7149 rosters).\\ 

\begin{table*}
\caption{\label{tab:table6}
Percentage of rosters (\%) associated with the workload metric that produced the highest fatigue (SP) or sleepiness (KSS) estimates within 28-day epochs for all rosters of 2023 (sample 4).}
\begin{ruledtabular}
\begin{tabular}{cccccccccc}
\multirow{2}{*}{Period} & \multirow{2}{*}{N} & Fatigue/Sleepiness & \multicolumn{7}{c}{Workload Metric}\\
& & Scale &$DT(h)$ & $N_{NS}$ & $N_{CREW}$ & $N_{DUTY}$ & $N_{SIT}$ & $N_{REST}$  & $N_{CREW}+N_{SIT}$\\
 \hline

\multirow{2}{*}{Jan-23} & \multirow{2}{*}{650} &SP	&25.7	&22.0	&23.2	&8.6	&N.A.	&3.2	&17.2\\
& & KSS	&29.5	&14.8	&18.5	&14.2	&12.8	& N.A.	&10.3\\
\multirow{2}{*}{Feb-23} & \multirow{2}{*}{508} & SP	&29.1	&13.0	&26.0	&11.6	&N.A.	&2.8	&17.5\\
& & KSS	&26.6	&9.8	&18.9	&17.5	&12.0	&N.A.	&15.2\\
\multirow{2}{*}{Mar-23} & \multirow{2}{*}{628} & SP	&23.9	&12.6	&24.8	&15.8	&N.A.	&4.3	&18.6\\
& & KSS	&24.2	&13.4	&18.0	&17.8	&11.3	&N.A.	&15.3\\
\multirow{2}{*}{Apr-23} & \multirow{2}{*}{628} & SP	&24.8	&14.2	&27.4	&9.2	&N.A.	&4.1	&20.2\\
& &KSS	&26.4	&9.6	&20.5	&14.5	&11.5	&N.A.	&17.5\\
\multirow{2}{*}{May-23} & \multirow{2}{*}{610} & SP	&25.2	&12.3	&28.2	&10.7	&N.A.	&4.3	&19.3\\
& & KSS	&22.6	&10.0	&22.0	&15.6	&12.8	&N.A.	&17.0\\
\multirow{2}{*}{Jun-23} & \multirow{2}{*}{614} & SP	&26.5	&15.1	&23.9	&11.4	&N.A.	&4.2	&18.7\\
& & KSS	&25.9	&9.6	&21.5	&15.6	&12.2	&N.A.	&15.1\\
\multirow{2}{*}{Jul-23} & \multirow{2}{*}{622} & SP	&22.2	&18.5	&28.1	&8.5	&N.A.	&3.4	&19.3\\
& & KSS	&23.5	&14.6	&22.5	&13.8	&10.3	&N.A.	&15.3\\
\multirow{2}{*}{Aug-23} & \multirow{2}{*}{658} & SP	&24.0	&17.0	&28.0	&7.8	&N.A.	&4.3	&18.9\\
& & KSS	&24.8	&11.6	&21.3	&16.4	&12.5	&N.A.	&13.4\\
\multirow{2}{*}{Sep-23} & \multirow{2}{*}{641} & SP	&26.8	&13.9	&27.3	&10.6	&N.A.	&3.3	&18.1\\
& & KSS	&24.0	&11.9	&21.5	&15.1	&11.7	&N.A.	&15.8\\
\multirow{2}{*}{Oct-23} & \multirow{2}{*}{580} & SP	&27.2	&15.7	&26.6	&9.8	&N.A.	&3.3	&17.4\\
& & KSS	&24.8	&14.1	&17.6	&14.5	&13.8	&N.A.	&15.2\\
\multirow{2}{*}{Nov-23} & \multirow{2}{*}{547} & SP	&28.7	&14.4	&24.5	&11.7	&N.A.	&4.6	&16.1\\
& & KSS	&27.1	&13.9	&17.6	&16.5	&11.7	&N.A.	&13.3\\
\multirow{2}{*}{Dec-23} & \multirow{2}{*}{464} & SP	&25.2	&18.3	&24.1	&11.6	&N.A.	&3.4	&17.2\\
& & KSS	&28.0	&14.0	&22.8	&11.4	&9.5	&N.A.	&14.2\\
\multirow{2}{*}{2023} & \multirow{2}{*}{7149} &SP	&25.7	&15.6	&26.1	&10.5	&N.A.	&3.8	&18.3\\
& & KSS	&25.6	&12.3	&20.2	&15.3	&11.9	&N.A.	&14.8\\

 \end{tabular}
\end{ruledtabular}
\end{table*}

\begin{table}
\caption{\label{tab:table7}
Percentage of all 168-hour loops within each 28-days epochs exceeding the proposed mitigations for $DT$, $N_{CREW}$ and $N_{CREW}+N_{SIT}$. Also shown the overall percentage of 2023. Details in the text.}
\begin{ruledtabular}
\begin{tabular}{ccccc}
\multirow{3}{*}{Period} & \multirow{3}{*}{N}  & \multicolumn{3}{c}{Mitigating Parameters}\\

& & $DT$ & $N_{CREW}$ & $N_{CREW}+N_{SIT}$\\
&&$>44$ h& $>15$& $>19$\\
 \hline

Jan-23 &650	&4.5	&2.4	&3.2\\
Feb-23 &508	&3.6	&2.5	&3.6\\
Mar-23 &628	&3.7	&2.8	&4.1\\
Apr-23 & 628	&5.1	&3.5	&4.3\\
May-23 & 610	&4.3	&3.7	&4.2\\
Jun-23 & 614	&4.0	&3.2	&4.5\\
Jul-23 & 622	&3.6	&4.0	&4.1\\
Aug-23 &657	&3.6	&3.1	&3.8\\
Sep-23 &641	&3.9	&3.6	&4.2\\
Oct-23 &580	&3.7	&2.5	&3.0\\
Nov-23 &547	&3.5	&2.7	&3.5\\
Dec-23 &464	&3.6	&2.7	&3.1\\
2023 &7149	&3.9	&3.1	&3.8\\
 \end{tabular}
\end{ruledtabular}
\end{table}

\section{Workload metrics within 28-days epochs for all rosters of 2023 (sample 4)}

The average values and standard errors of the workload metrics computed within 28-days epochs for all rosters of 2023 are presented in Table \ref{tab:table8}. 

 \begin{table*}
\caption{\label{tab:table8}
Averages and standard errors (SE) of the workload metrics computed within 28-day epochs for all 7149 rosters of 2023 (sample 4).}
\begin{ruledtabular}
\begin{tabular}{cccccccccccccc}
Period & N & Parameter & $N_{NS}$ & $N_{CNS}$&$N_{ES}$ &$DT(h)$ & $N_{CREW}$ & $N_{WOCL}$ & $N_{SIT}$ & $N_{REST}$ & $N_{DUTY}$ & $N_{CREW}+N_{SIT}$\\
 \hline

\multirow{2}{*}{Jan-23} & \multirow{2}{*}{650} & Average & 6.53 &2.28	&1.75 &	109.3	&31.0 &	4.50&	7.69	&5.33	&3.93	&38.7\\
& & SE &0.10 &0.06	&0.06&	0.8	&0.3	&0.12	&0.14	&0.09	&0.08	&0.5\\
\multirow{2}{*}{Feb-23} & \multirow{2}{*}{508} & Average &5.50&1.80	&1.96	&102.2	&29.4	&3.70&	7.87&	4.71&	3.81&	37.3\\
& & SE &0.10&0.06	&0.08&	1.1&	0.4&	0.12&	0.16&	0.10&	0.08&	0.5\\
\multirow{2}{*}{Mar-23} & \multirow{2}{*}{628} & Average &5.60&1.74	&1.85&	102.5&	28.9&	3.84&	7.94&	5.06&	3.70&	36.8\\
& & SE &0.10&0.05	&0.06&	1.0&	0.4&	0.12&	0.15&	0.10&	0.08&	0.5\\
\multirow{2}{*}{Apr-23} & \multirow{2}{*}{628} & Average &5.78&1.79	&1.71&	107.5&	30.2&	3.86&	7.88&	5.44&	4.00&	38.1\\
& & SE &0.10&0.05	&0.06&	1.0&	0.4&	0.12&	0.15&	0.11&	0.09&	0.5\\
\multirow{2}{*}{May-23} & \multirow{2}{*}{610} & Average &5.78&1.81	&1.68&	104.9&	30.0&	3.64&	7.66&	5.36&	3.78&	37.6\\
& & SE &0.10&0.05	&0.06&	1.0&	0.4&	0.11&	0.14&	0.10&	0.08&	0.5\\
\multirow{2}{*}{Jun-23} & \multirow{2}{*}{614} & Average &5.97&1.84	&1.77&	105.3&	30.0&	3.70&	7.91&	5.37&	3.82&	37.9\\
& & SE &0.10&0.05	&0.06&	1.0&	0.4&	0.11&	0.15&	0.11&	0.08&	0.5\\
\multirow{2}{*}{Jul-23} & \multirow{2}{*}{622} & Average &6.35&2.04	&1.75&	106.9&	31.8&	4.40&	7.57&	5.44&	3.80&	39.3\\
& & SE &0.10&0.05	&0.06&	0.9&	0.4&	0.13&	0.14&	0.10&	0.08&	0.5\\
\multirow{2}{*}{Aug-23} & \multirow{2}{*}{658} & Average &5.91&1.87	&1.63&	102.2&	29.3&	3.77&	7.20&	5.09&	3.74&	36.5\\
& & SE &0.10&0.05	&0.06&	1.0&	0.4&	0.11&	0.14&	0.10&	0.08&	0.5\\
\multirow{2}{*}{Sep-23} & \multirow{2}{*}{641} & Average &5.81&1.83	&1.66&	102.5&	29.3&	3.73&	7.29&	5.02&	3.79&	36.6\\
& & SE &0.09&0.05	&0.06&	1.0&	0.4&	0.12&	0.13&	0.10&	0.08&	0.5\\
\multirow{2}{*}{Oct-23} & \multirow{2}{*}{580} & Average &5.62&1.77	&1.82&	103.0&	28.7&	3.84&	7.16&	4.94&	3.82&	35.9\\
& & SE &0.10&0.05	&0.07&	1.1&	0.4&	0.12&	0.14&	0.11&	0.08&	0.5\\
\multirow{2}{*}{Nov-23} & \multirow{2}{*}{547} & Average &5.22&1.61	&1.85&	97.8&	27.3&	3.37&	7.14&	4.78&	3.35&	34.4\\
& & SE &0.11&0.05	&0.07&	1.1&	0.4&	0.13&	0.15&	0.10&	0.08&	0.5\\
\multirow{2}{*}{Dec-23} & \multirow{2}{*}{464} & Average &5.75&1.91	&2.00&	101.0&	28.3&	4.05&	7.10&	4.77&	3.30&	35.4\\
& & SE &0.12&0.06	&0.08&	1.1&	0.5&	0.15&	0.16&	0.11&	0.09&	0.6\\

 \end{tabular}
\end{ruledtabular}
\end{table*}

\section{On-line data-tables}
The results presented in Figures 1 through 5 will be available in electronic format after the publication of the final version of this work.

\FloatBarrier

\bibliography{Manuscript_arxiv}

\end{document}